# Large antiferromagnetic fluctuation enhancement of the thermopower at a critical doping in magnetic semimetal $Cr_{1+\delta}Te_2$

*Atwa Mohamed[1], Yuita Fujisawa[1], Takatsugu Onishi[1], Markel Pardo-Almanza[1], Mathieu Couillard[1],*

*Keita Harada[1], Tsunehiro Takeuch[2], Yoshinori Okada[1]*

[1]Quantum Materials Science Unit, Okinawa Institute of Science and Technology (OIST),

Okinawa 904-0495, Japan

[2]Research Center for Smart Energy Technology, Toyota Technological Institute,

Nagoya 468-8511, Japan;

**Abstract**

$Cr_{1+\delta}Te_2$ is a self-intercalated transition metal dichalcogenide that hosts tunable electronic filling and magnetism in its semimetallic band structure. Recent angle-resolved photoemission spectroscopy (ARPES) studies have unveiled a systematic shift in this semimetallic band structure relative to the chemical potential with increased Cr doping. This report presents the temperature and magnetic field dependence of the longitudinal thermopower $S_{xx}$ for different $Cr_{1+\delta}Te_2$ compositions. We show that as doping increases, the sign of $S_{xx}$ changes from positive to negative at the critical doping level of $\delta \sim 0.5$. This observed doping-dependent trend in the thermopower is consistent with the evolution of the semimetallic band structure from ARPES. Importantly, an anomalous enhancement of the thermoelectric response near $T_C$ is also observed around $\delta \sim 0.5$. Combining information from magnetometry and ARPES measurements, existence of the critical nature of the doping level $\delta_c$ ($\sim 0.5$) is unveiled in magnetic semimetal $Cr_{1+\delta}Te_2$, where antiferromagnetic fluctuation and near-Fermi-energy pseudogap formation play a potential vital role in enhancing thermoelectric energy conversion.

**Introduction**

The thermoelectric effect enables the conversion of thermal energy to electricity[1]. Asymmetry in the energy-dependent electronic conductivity between the hot and cold sides of a metallic conductor generates a driving force for a net flow of charge entropy. Whatever the microscopic origins of this entropy are, if they are coupled to charge carriers through inelastic scattering, one can imagine various intriguing ways to modulate thermoelectric effects. Manipulation of the magnetic degree of freedom is another attractive route to controlling the thermoelectric effect. For example, while a consensus has yet to be reached, a large spin entropy and anomalous electronic states near $E_F$ have been proposed to explain the giant Seebeck effect of cobalt oxides[2,3,4,5]. Efforts to simultaneously engineer the heat, spin, and charge degrees of freedom have resulted in the field of spin caloritronics, where spin currents control and enhance thermoelectric phenomena[6,7,8,9,10]. Magnetic fluctuation driven enhancement of the thermopower has been reported, for example, in a Heusler compound[12]. Antiferromagnetism enhanced thermoelectricity has garnered considerable interest in from both a fundamental and practical perspective[13,14]. Research on the interaction between improved thermoelectric response and Berry curvature physics is also ongoing[15,16,17]. Although these recent explorations of the exotic interplay between magnetism and the thermoelectric response are promising, a microscopic understanding of the coupling between magnetism and thermoelectricity is still in its infancy. To promote understanding of the interaction between spins and charges in thermoelectric materials, ideal materials to focus on are those with widely tunable electronic band filling and deeply controllable magnetic structure.

Transition-metal dichalcogenides (TMDs) are a unique platform for tuning electronic and magnetic properties in low dimension. The atomic layers of TMDs are weakly coupled through the van der Waals (vDW) force, creating an opportunity to chemically tune their physical properties by intercalating native atoms into the vDW gaps[18]. Among the magnetic TMDs, $Cr_{1+\delta}Te_2$ (**Fig. 1a**) is a promising self-intercalated ferromagnet with widely-tunable electronic and magnetic properties[19,20,21,22,23,24,25,26,27,28,29,30,31,32,33,34,35,36,37]. Our recent efforts have yielded a unique recipe for the growth of epitaxial thin films of $Cr_{1+\delta}Te_2$ over a wide compositional range[38]. Based on the previously reported phase diagram of this system, a Curie temperature ($T_c$) beyond room temperature is achieved as the intercalated Cr ($\delta$) increases (**Fig. 1b left axis**). Furthermore, the effective magnetic anisotropy energy ($K_{eff}$), defined as the difference in energy to align spins along the in-plane ($K_{in}$) or out-of-plane ($K_{out}$) directions, shows a gradual change from positive to negative values as $\delta$ increases (**Fig. 1b right axis**). Density functional theory (DFT) calculations have suggested an inherent magnetic frustration from competing magnetic interactions, which plays a role in modulating $K_{eff}$ in this system. This $K_{eff}$ modulation yielding enhanced magnetic fluctuation at $K_{eff} \sim 0$ has been speculated in literature[38].

Recently, the doping-dependent electronic structure of $Cr_{1+\delta}Te_2$ has also been reported. Using in-situ angle-resolved photoemission spectroscopy (ARPES), a systematic energy shift of the semimetallic band portion was directly revealed around the zone boundary relative to the chemical potential (**Figs. 1c and d**)[39]. ARPES measurements at 14 K (well below $T_C$) have yielded a characteristic energy ($E_0$) for the semimetallic band portion around the $\overline{M}$ point (**Fig. 1d**). The Fermi energy $E_F$ was shown to approach $E_0$ near a critical doping $\delta_c$ ~ 0.5. Alongside a metallic hole band near the $\overline{\Gamma}$ point at all doping levels, the formation of a pseudogap in the density of states (DOS) was predicted to occur around this critical composition. These ARPES findings prompted our current investigation of the doping-dependent longitudinal thermopower ($S_{xx}$), as its sign and absolute value should be sensitive probes of the existence of the pseudogap and its energy relative to the chemical potential. Another worthwhile motivation for measuring $S_{xx}$ is to pursue any magnetism-driven enhancements to the thermopower that $Cr_{1+\delta}Te_2$ can be expected to host, being both magnetically and electronically tunable. However, the doping evolution of the thermoelectric response in the absence and presence of magnetic fields has not been investigated in $Cr_{1+\delta}Te_2$.

We report on the systematic measurement of $S_{xx}$ and longitudinal electrical resistivity ($\rho_{xx}$) in magnetic semimetal $Cr_{1+\delta}Te_2$. We discuss the doping, temperature, and magnetic field evolution of $S_{xx}$ with respect to the underlying band structure and magnetism. As the most significant finding in this study, we show that at the critical doping level of $\delta_c \sim 0.5$, a possible interaction between magnetic fluctuations and the formation of an anomalous electronic state at the chemical potential cooperatively leads to an enhancement of the thermoelectric properties in $Cr_{1+\delta}Te_2$.

## Methods

### Sample Preparation

The (001) oriented epitaxial $Cr_{1+\delta}Te_2$ films used in this work were grown on $Al_2O_3$ (0001) substrates with a molecular beam epitaxy (MBE) system through a two-step process that involved film deposition followed by post-deposition annealing in situ as described in our previous work. Following the same detailed characterization methods performed previously, the fraction of intercalated $Cr(\delta)$ was determined by combining energy-dispersive X-ray spectroscopy (EDS) of the elemental ratios and X-ray diffraction (XRD) estimates of the lattice constants. The thickness of all samples grown in this study was approximately 80 nm.

### Transport Measurements

$S_{xx}$ and $\rho_{xx}$ were measured simultaneously using a Quantum Design PPMS® DynaCool system combined with a custom-built sample stage and electronics set-up. The electrode configuration is shown in **Fig. 2a**. In this study, technical limitations in our measurement setup restrict our $S_{xx}$ measurements between 80 and 380 K. The lower temperature limit mainly represents a threshold below which the $Al_2O_3$ substrate becomes too thermally conductive to maintain a sufficient temperature gradient and determine $S_{xx}$ across the sample accurately. The vertical dotted lines in **Fig. 1b** indicate the doping levels studied in this report on the previously reported phase diagram. In this report, all magnetic fields were applied along the out-of-plane direction, as shown schematically in **Fig. 2a**.

## Results and Discussion

### Doping and temperature dependence of $\rho_{xx}$ and $S_{xx}$ in $B=0$

The temperature and doping dependence of $\rho_{xx}$ and $S_{xx}$ are shown in **Figs. 2b** and **c**, respectively. Regardless of the detailed microscopic origins, suppression of electron-spin scattering can be expected to occur in magnetically-ordered states compared to paramagnetic states. Therefore, $T_C$ at each doping level is estimated from the position of the kink in the $\rho_{xx}(T)$ curves, as indicated by the arrows in **Figs. 2b** and **c**. The estimated $T_C$ values are consistent with those determined previously[38]. The doping dependence of $S_{xx}(T)$ also exhibits several characteristic behaviors. The samples with $\delta = 0.34$ and $\delta = 0.4$ show positive values with a nearly linear $T$ dependence (left two panels in **Fig. 2c**). In contrast, $S_{xx}$ for $\delta = 0.50$ and $\delta = 0.54$ (middle panel in **Fig. 2c**) show negative values with a kink around $T_C$. Notably, such a kink structure around $T_C$ is nearly absent in the highest doped sample, $\delta = 0.68$ (right panel in **Fig. 2c**). Hereafter, we focus on $S_{xx}$, as the essential trends in $\rho_{xx}(T)$ are qualitatively similar for all doping levels. In the following sections, we elaborate on the doping-dependent sign change in $S_{xx}(T)$, followed by a discussion of the electronic and magnetic origins of the kink around $T_C$ in the $S_{xx}(T)$ curves.

**Mott formula for $S_{xx}$**

The simplest model for the temperature dependence of thermopower $S_{xx}$ in metals is the so-called Mott relation.

$$S_{xx} = \frac{\pi^2 k_B^2 T}{3e} \times \left\{\frac{d\ln \sigma_{xx}(E)}{dE}\right\}_{E=E_F} = \frac{\pi^2 k_B^2 T}{3e}\left[\frac{1}{N(E_F)}\left\{-\frac{dN(E)}{dE}\right\}_{E=E_F} + \frac{1}{\tau(E_F)}\left\{\frac{d\tau(E)}{dE}\right\}_{E=E_F}\right] \quad (1)$$

Here, the spectral conductivity $\sigma_{xx}$ is proportional to the density of states $N(E)$ and the scattering rate $\tau(E)$, based on the relaxation time approximation from Boltzmann transport theory. By assuming an energy-independent scattering rate $\tau$ in Equation (1), $S_{xx}$ can be expressed as:

$$S_{xx} \approx \frac{\pi^2 k_B^2 T}{3e}\left[\frac{1}{N(E_F)}\left\{-\frac{dN(E)}{dE}\right\}_{E=E_F}\right] \quad (2)$$

The above relation makes it evident that the sign of $S_{xx}$ corresponds to that of $-dN(E)/dE_{E=E_F}$. In this case, if the chemical potential exists in a hole-like band ($-dN/dE > 0$ at $E_F$), the sign and slope of $S_{xx}(T)$ become correspondingly positive, while if the chemical potential exists in an electron-like band ($-dN/dE < 0$ at $E_F$), the sign and slope of $S_{xx}(T)$ become negative.

**Relation between $S_{xx}(\delta)$ and band structure**

To uncover the connection between the sign of $S_{xx}$ and the corresponding band structure at each Cr($\delta$), we plot the doping dependence of $S_{xx}$ at 350 K and its correlation with the doping evolution of the semimetallic band portion at the $\bar{M}$ point (**Fig. 2d**). For a fair comparison, 350 K is chosen to be above $T_C$ for all doping levels. A sign change in $S_{xx}$ at 350 K is evident as $\delta$ increases. Considering the semimetallic band around the $\bar{M}$ point (**Fig. 1d**), $E_F(\delta) - E_0 > 0$ leads to $-dN/dE > 0$ and consequently to a positive sign in $S_{xx}$. On the other hand, $E_F(\delta) - E_0 < 0$ leads to $-dN/dE < 0$, corresponding to a negative sign in $S_{xx}$. Admittedly, this is a simplified picture that ignores the existence of bands other than the semimetallic band around the $\bar{M}$ point. Moreover, this simplified picture deliberately excludes the nuance of a $k_z$ dispersion relying solely on the band dispersion around the $\bar{M}$ point probed using the single-photon energy available in our ARPES[38]. Nevertheless, a correlation can clearly be seen between the sign of $E_F(\delta) - E_0$ and $S_{xx}(\delta)$. This correlation implies that the previously observed ARPES band around the $\bar{M}$ point in **Fig. 1c** governs the doping-dependent behavior of the thermoelectric properties in this system. This simple picture is particularly justified at higher temperatures such as those examined in this study, as the near-$E_F$ fine band structure beyond considerations of a simple picture in **Fig. 1c** is thermally smeared out even if it exists. Also, note that the doping dependence of normal Hall coefficient does not show sign change with doping $\delta$ in a previous report[38]. However, this is not contradictory. While the normal Hall effect is more sensitive to electronic state anisotropy in momentum space at $E_F$[40], $S_{xx}$ is sensitive to electronic state anisotropy along energy axis relative to $E_F$.

**Magnetic fluctuation driven enhancement of $S_{xx}$**

Next, we discuss the origin of the kink in the $S_{xx}(T)$ curves across $T_C$. Based on Equation (1), $S_{xx}$ can be modulated by changes to both $N(E)$ and $\tau(E)$ by the magnetic phase transition. Before discussing these quantities, we first elaborate on the nature of magnetic fluctuations above $T_C$. Focusing on the magnetic field dependence of $S_{xx}$ is a rational method of investigating the influence that magnetic fluctuations have on the thermoelectric response[41]. **Fig. 3a** shows $S_{xx}$ in an external magnetic field $B = 0$ (filled symbols) and 9 T (empty symbols). To quantify the magnetic field-dependent contribution, $|\Delta S_B(T)|=|S_{xx}(T)_{B = 9\,T} - S_{xx}(T)_{B = 0\,T}|$ is defined (see hatched area in **Fig. 3a**). Here, $B = 9$ T is chosen to be large enough compared to saturation field to completely suppress magnetic fluctuations in a wide temperature region both above and below $T_C$[38]. As the application of a magnetic field 9 T suppresses magnetic fluctuations, the quantity $|\Delta S_B(T)|$ can be considered a measure of the magnetic fluctuation-related contributions to $S_{xx}$ in the absence of a magnetic field (**Fig. 2**). Such method was previously employed in studying the magnetic fluctuation enhancement of the thermopower in Heusler alloys[12]. In our case, while $|\Delta S_B(T)|$ for two doping levels $\delta = 0.34$ and 0.40 are negligible, $|\Delta S_B(T)|$ becomes prominent for other three doping levels $\delta = 0.5$, 0.54, and 0.68. This observation indicates that $S_{xx}$ responds more sensitively to an external magnetic field around $\delta = 0.5$, which results in a maximum $|\Delta S_B(T)|$ at $T_C$ around this doping level, as opposed to a monotonic increase of $|\Delta S_B(T_C)|$ with $\delta$. Hereafter, we discuss the intertwined microscopic nature of the magnetic fluctuations and the enhancement in $S_{xx}$, focusing on the critical doping level $\delta \sim 0.5$, which we designate $\delta_c$.

**Spin Fluctuation vs. Magnon Pictures around $T_C$**

We begin by discussing the microscopic picture of the fluctuating magnetism around $T_C$. The conventional mean-field picture for explaining magnetism in metals is known as the Stoner model. In this model, spin-degenerate bands are split into majority and minority spin bands separated by an exchange energy that is associated with the energetic cost of transitioning from paramagnetic states ($T > T_C$) to magnetically ordered states ($T < T_C$)[42]. An abrupt change in the band structure occurs near $T_C$ per this interpretation. However, this conventional Stoner model can be excluded as the origin of the magnetic behavior in our system, as we observe signature of magnetic fluctuation far above $T_C$ (**Fig. 3b-c**). As an alternative, we invoke the two pictures of the microscopic origins of magnetic fluctuations. The first picture is spin fluctuation theory, which has succeeded in explaining the physical properties of itinerant magnetism in various materials[43]. The second picture relies on the collective propagations of magnetic spin precessions, known as magnons, which emerge in localized magnetically ordered states. Spin fluctuations in itinerant electron systems predominantly influence the thermodynamic properties of weakly or nearly ferromagnetic metals, and such spin fluctuations can survive well above $T_C$. Although the conventional formalism of magnons supports their dominant existence only below the Curie and Néel temperatures ($T_C$ and $T_N$) of localized, magnetically-ordered systems[44], magnon excitations have been shown to persist above $T_C$ in the so-called paramagnon regime[45].

**Logarithmic temperature dependence $T^n$**

To discern the nature of the magnetic fluctuations enhancing the thermopower, examining the exponents of the logarithmic thermopower temperature dependence has been recognized as a fruitful approach [12]. The exponent of $|\Delta S_B(T)|$ up to $T_C$ was used to clarify the origin of the magnetic fluctuation enhanced thermopower. This approach is underpinned by microscopic models of the magnetic fluctuation contributions to the specific heat ($C_{mag}$), which are reflected in the thermopower temperature dependence below the transition temperature [12]. While an itinerant spin fluctuation picture does not support the existence of a well-defined logarithmic temperature dependence[46], the magnon picture does support a logarithmic temperature dependence[47]. Based on the hydrodynamic theory of magnon-electron drag[44] $|\Delta S_{xx}|$ can be expressed as:

$$|\Delta S_{xx}| = \frac{2}{3} \frac{C_{mag}}{ne} \frac{1}{1+\tau_{em}/\tau_m}.$$

Here, $\tau_{em}$ is the electron-magnon scattering time and $\tau_m$ is the total magnon scattering time for all magnon scattering events (i.e., electron-magnon, phonon-magnon, magnon-magnon) respectively. Note that $\tau_{em} \geq \tau_m$ holds by definition. If $\tau_{em} \gg \tau_m$ holds (indicating weak electron-magnon coupling), the drag contribution in $S_{xx}$ becomes negligible. On the other hand, if $\tau_{em} \sim \tau_m$ holds (indicating strong electron-magnon coupling), then the temperature dependence of drag contribution in $S_{xx}(T)$ is governed by $C_{mag}(T)$. Importantly, based on magnon picture, $C_{mag}(T)$ shows logarithmic temperature dependence $T^n$, and $C_{mag}$ of FM and AFM magnons are known to exhibit $n \sim 1.5$ and $n \sim 3$, respectively[47]. In our case, $|\Delta S_B|$ is expected to reflect $|\Delta S_{xx}|$, and it is a reasonable simplification not to consider a phonon drag contribution since our $|\Delta S_B|$ is dominated nearby magnetic phase transition temperature $T_C$.

**Existence of exponent n~3**

To check for the existence of a logarithmic temperature dependence of $|\Delta S_B(T)|$ near $T_C$, we first show $\log(\Delta S_B)$ vs. $\log(T)$ plot in **Fig. 3b-d**. From this plot, the existence of a logarithmic temperature dependence of $|\Delta S_B(T)|$ near $T_C$ is visually discernable from the linear dependence. Indeed, around $T_C$, a clear linearity can be recognized from $\log(\Delta S_B)$ vs. $\log(T)$ plots of the three doping levels $\delta = 0.5$, 0.54, and 0.68. From the fitting of the temperature region between $T_C$ and ~ 100 K below $T_C$ at each doping level, the exponents $n$ for three doping levels $\delta = 0.5$, 0.54, and 0.68 are 3.2, 2.8, and 2.7, respectively (see lined in **Fig. 3b-d**). We also show the temperature dependence of the slope as defined by following equation $n(T) = \partial\{\log(\Delta S_B) - \log(T)\}/\partial \log(T)$. From the plot shown in **Fig. 3e**, this slope $n$ is almost constant near $T_C$, and the averaged estimated value of n at each of the three doping levels in this temperature region is around $n \sim 3$. Phenomenologically, our observation of an $n = 3$ at three different doping levels of $\delta \geq \delta_C$ strongly implies the existence of AF magnon drag with $\tau_{em} \sim \tau_m$, rather than an $n \sim 1.5$ based FM magnon picture. While the drag effect is seen mainly at very low temperatures where electron-boson coupling is strong, our magnon case shows a dominant drag contribution around $T_C$. The potential reason leading to $\tau_{em} \sim \tau_m$ is our nature of magnetism in our system, which at the cusp of itinerant and localized magnetism. Although quantifying $\tau_{em}$ and $\tau_m$ are generally challenging, similar magnon contributions have been observed near $T_C$ in several magnetic materials[48,49,50].

**Nature of critical doping**

Our most intriguing finding beyond the existing literature on the Cr-Te system[27,51,52,53,54] is the existence of a critical doping ($\delta_c$). From a previous DFT study, FM interactions couple Cr atoms along the in-plane direction on one sublattice, while AFM interactions couple Cr atoms along the out-of-plane direction on the other sublattice, in the case of CrTe ($\delta = 1$) with two clearly differentiated magnetic Cr sublattices[38]. Even in the critical case of $\delta_c \sim 0.5$, it is natural to assume the coexistence of FM and AFM interactions, as shown in **Fig. 4c**. Intriguingly, the experimental realization of $K_{eff} \sim 0$ around $\delta_c$ is expected to reflect a particularly unique situation where the competition between multiple magnetic interactions and a correspondingly large frustration leads to enhanced magnetic fluctuations[38]. Notably, the dominant exponent $n \sim 3$ around critical doping (**Fig. 3**) is consistent with expected crucial role of AFM based fluctuations.

**Electronic structure perspectives**

We speculate that the magnetic fluctuation driven enhancement of $S_{xx}$ around $\delta_c$ is also linked to the anomaly in the electronic state around this critical doping level. This notion is supported by the fact that $E_0-E_F=0$ is realized around $\delta_c \sim 0.5$. As indicated previously, while a metallic hole band occurs near the $\bar{\Gamma}$ point at all doping levels, a pseudogap in the density of states at $E_F$ occurs exclusively at $\delta_c$ (see **Fig. 1c** and **d**). Therefore, we suspect that enhanced magnetic fluctuations alongside the pseudogap formation at $E_F$ in the $\delta_c$ sample cooperatively drive the observed enhancement in the zero-field thermoelectric energy conversion around $T_C$. There are two paradigms by which to interpret such a cooperation. The first relies solely on the experimentally observed band structure around the $\bar{M}$ point from ARPES, disregarding the energy dependence of the scattering rate $\tau(E)$. This picture is justified by the factor $1/N(E)$ in an application of Equation (2) to our observations, since the position $E_0 - E_F = 0$ is expected to lead a dip in $N(E)$ (i.e., a pseudogap). A second and more holistic paradigm would be to consider the asymmetry in $\tau(E)$ relative to $E_F$, in addition to expected pseudogap in $N(E)$. For instance, such asymmetric $\tau(E)$ can be realized if a characteristic energy for an electronic state sensitive to the spin orientation and its fluctuations exists away from $E_F$. Although modeling of the microscopic underpinnings of these observations in $S_{xx}$ is anticipated as a future undertaking, it can be expected that a significant spin-orbit coupling effect arising from the presence of the heavy element Te plays a role in bridging between the spin and charge degrees of freedom near $E_F$ in $Cr_{1+\delta}Te_2$.

**Summary**

A systematic investigation of the doping, temperature, and magnetic field dependence of the longitudinal thermoelectric response $S_{xx}$ is presented in the electronically/magnetically tunable semimetal $Cr_{1+\delta}Te_2$. We show signatures of magnetic fluctuation-driven enhancement of longitudinal thermoelectric response $S_{xx}$ around a critical doping level $\delta_c$~0.5, where antiferromagnetic fluctuations and near-Fermi-energy pseudogap play a potential vital role in enhancing thermoelectric energy conversion.

We emphasize that detection of a magnetically modulated thermoelectric signal necessitates coupling between the magnetic and charge degrees of freedom. While a solid understanding of the underlying electronic states is always crucial in interpreting such charge-spin coupling, direct spectroscopic evidence of the electronic structure has been lacking from most studies in magnetic/metallic thermoelectric materials so far. As such, this study, which bridges between momentum space electronic states and thermoelectric effects in the tunable magnetic semimetal $Cr_{1+\delta}Te_2$, provides valuable clarifying information regarding the interplay between magnetism and thermoelectricity. Finally, the fact that $Cr_{1+\delta}Te_2$ has been identified as an intriguing material platform that hosts Berry curvature physics in real and momentum spaces also suggests the possibility of exotic intertwined effects between the anomalous thermoelectric response and Berry curvature physics to be pursued in future investigations of this material, including study of transverse thermoelectric effect.

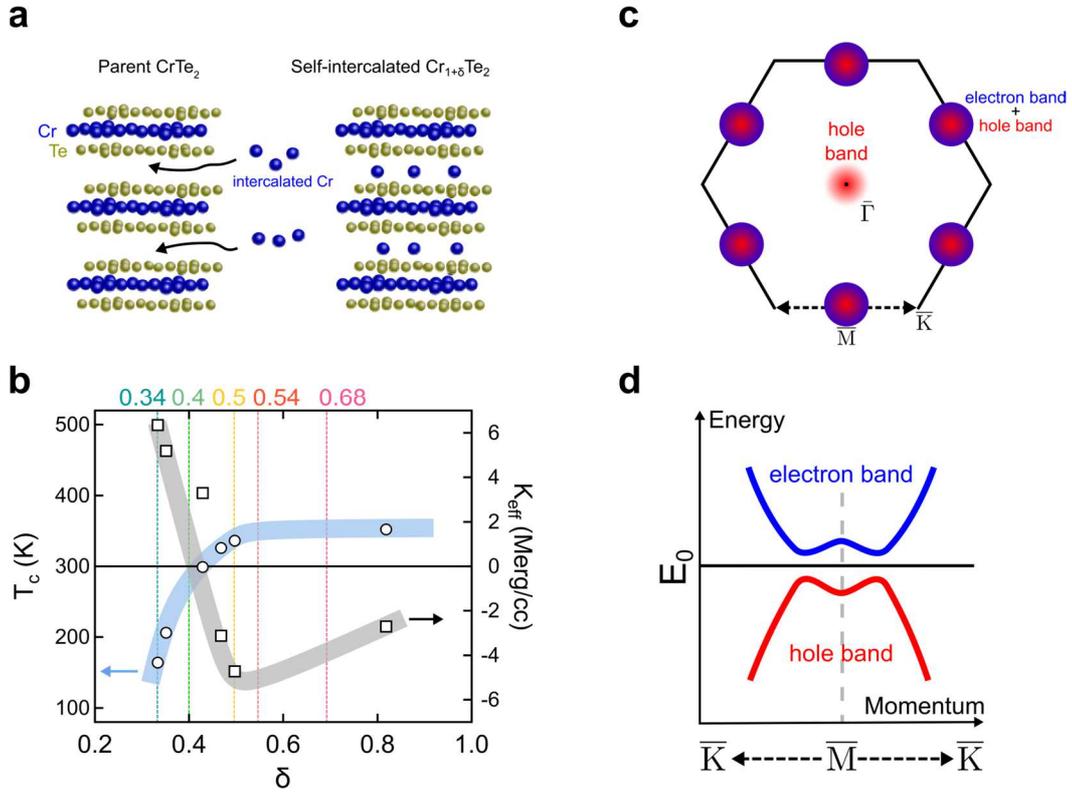

**Fig. 1** Magnetic and electronic tunability of $Cr_{1+\delta}Te_2$. **a** Schematic for self-intercalation of Cr atoms in the parent $CrTe_2$. **b** Doping evolution of the Curie temperature ($T_C$) and magnetic anisotropy energy ($K_{eff}$) as previously determined from [38]. The colored lines indicate the compositions investigated in this study. **c** Schematic of the *k*-space electronic structure determined from ARPES studies on $Cr_{1+\delta}Te_2$ [39]. Around $\bar{\Gamma}$, the hole band dominates near the Fermi energy. However, the coexistence of electron and hole bands around $\bar{M}$ points constitute a semimetallic band portion. **d** The schematic semimetallic band portion around $\bar{M}$ from **c**. The characteristic energy ($E_0$) is the charge neutral point of the semimetallic band portion corresponding to a pseudogap.

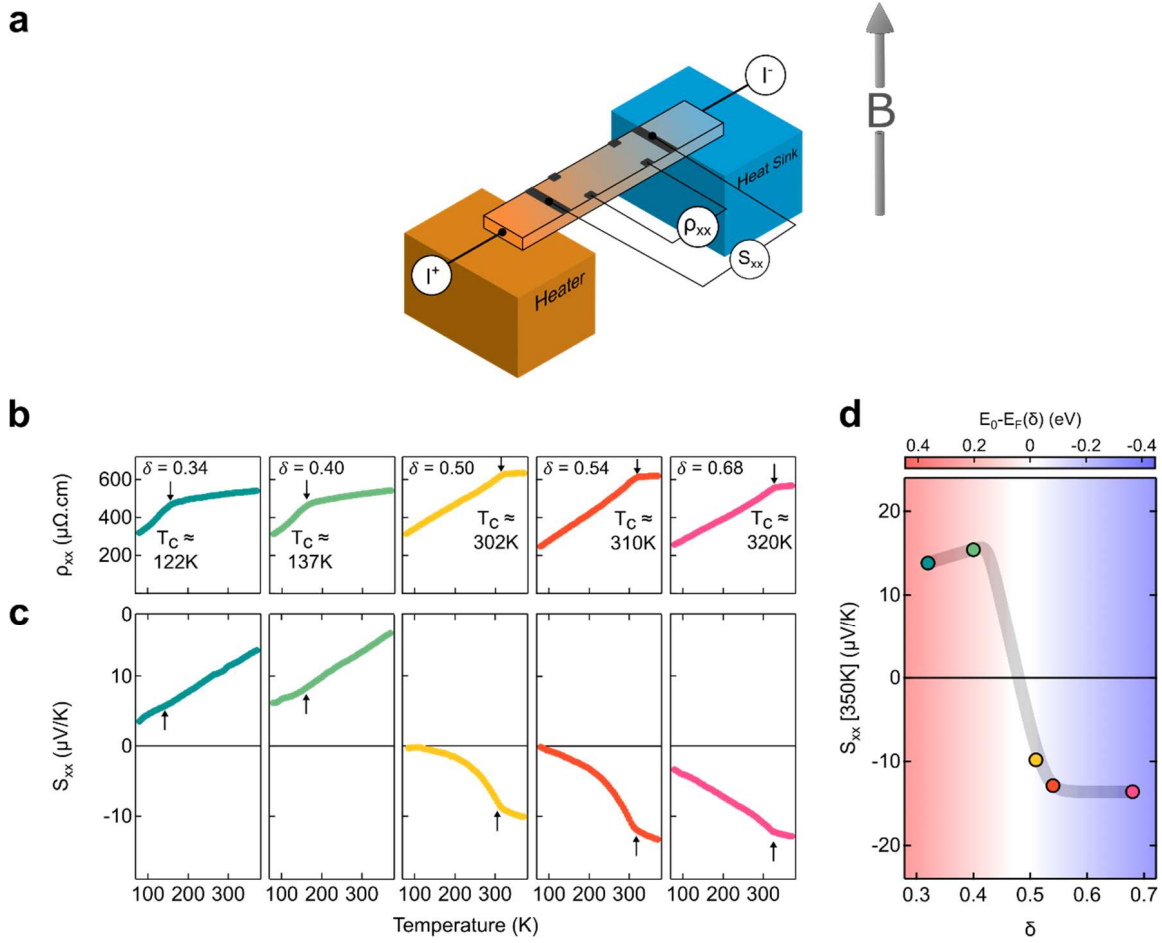

**Fig. 2** Evolution of the transport properties in (001) oriented epitaxial $Cr_{1+\delta}Te_2$ films grown on $Al_2O_3$ (0001) substrates. **a** Electrode configuration for simultaneous measurement of longitudinal resistivity ($\rho_{xx}$) and thermopower ($S_{xx}$). The *B*-field is applied parallel to $Cr_{1+\delta}Te_2(001)$ direction. **b** Temperature and doping evolution of $\rho_{xx}$. $T_C$ is estimated from the kinks in $\rho_{xx}(T)$. **c** Temperature dependence of $S_{xx}$. Arrows denote $T_C$ from **b**. **d** Doping evolution of $S_{xx}$ at 350 K with $E_0$-$E_F(\delta)$ from [39] overlayed as a gradient showing a clear sign crossover in both quantities around $\delta = 0.5$.

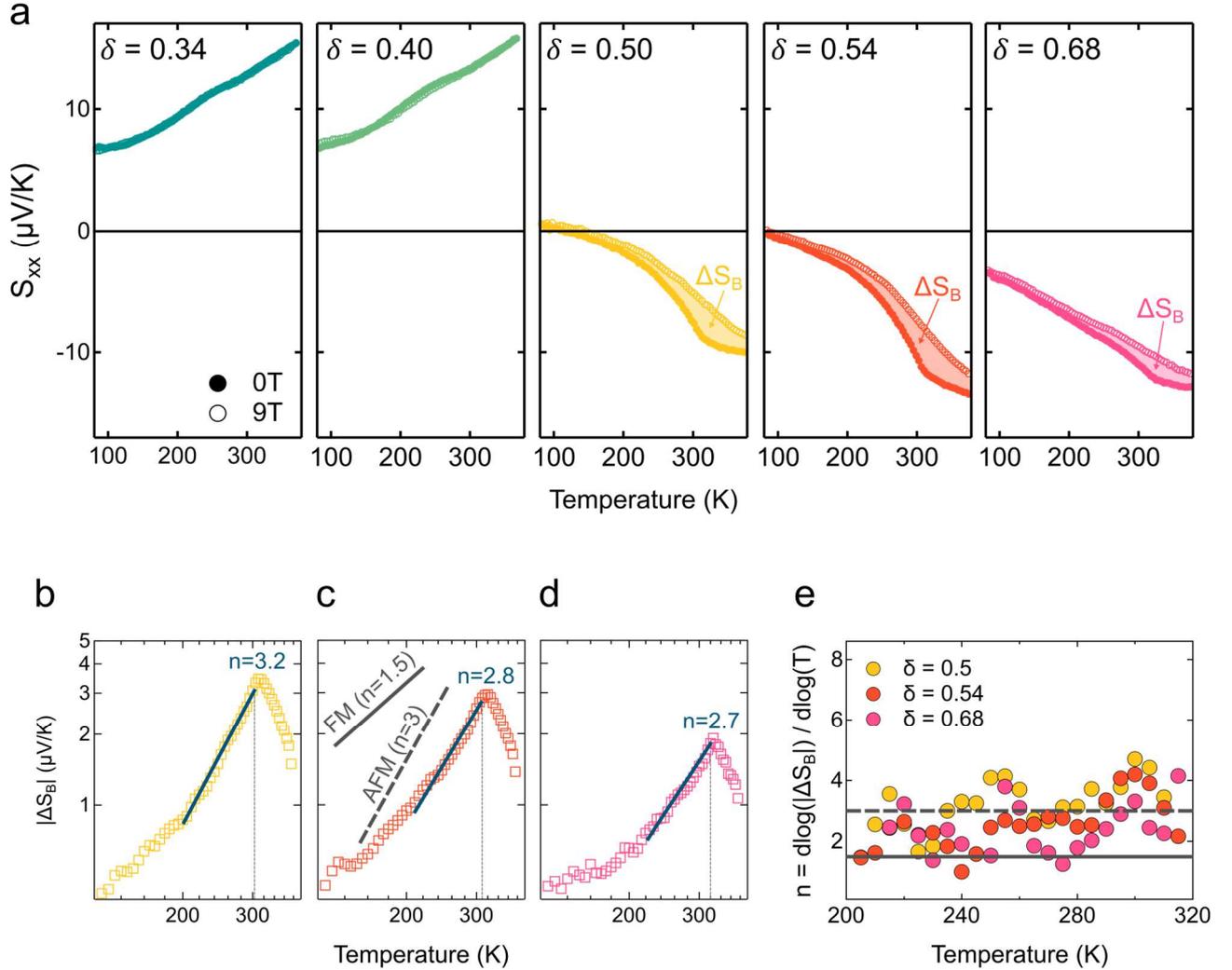

**Fig. 3** Doping evolution of the magneto thermopower. **a** Comparison between the thermopowers at 0 and 9 T for the five doping levels. A significant difference between $S_{xx}(0\,T)$ and $S_{xx}(9\,T)$ can be seen in the $\delta = 0.5$ and $\delta = 0.54$, indicated by the shaded regions and defined as $\Delta S_B$. **b-d** log-log plots of $|\Delta S_B|=|S_{0\,T} - S_{9\,T}|$ for the $\delta = 0.5$, 0.54 and 0.68 samples. $T_C$ is indicated at each doping level by the thin grey line. Power fits of the linear regions (between $T_C$ and ~100 K below $T_C$) on the log-log plot are indicated by the solid blue lines with $T^n$ exponents indicated. The slopes on the log-log plots corresponding to FM and AFM magnons are indicated by the solid and dashed grey lines in **c**. **e** The temperature dependence of the fitting exponent $n$ determined by $n(T) = \partial\{\log(\Delta S_B) - \log(T)\}/\partial \log(T)$ for the three doping levels ($\delta = 0.5$, $\delta = 0.54$ and $\delta = 0.68$). The solid and broken lines in **c** and **e** are $n$ for FM and AFM cases.

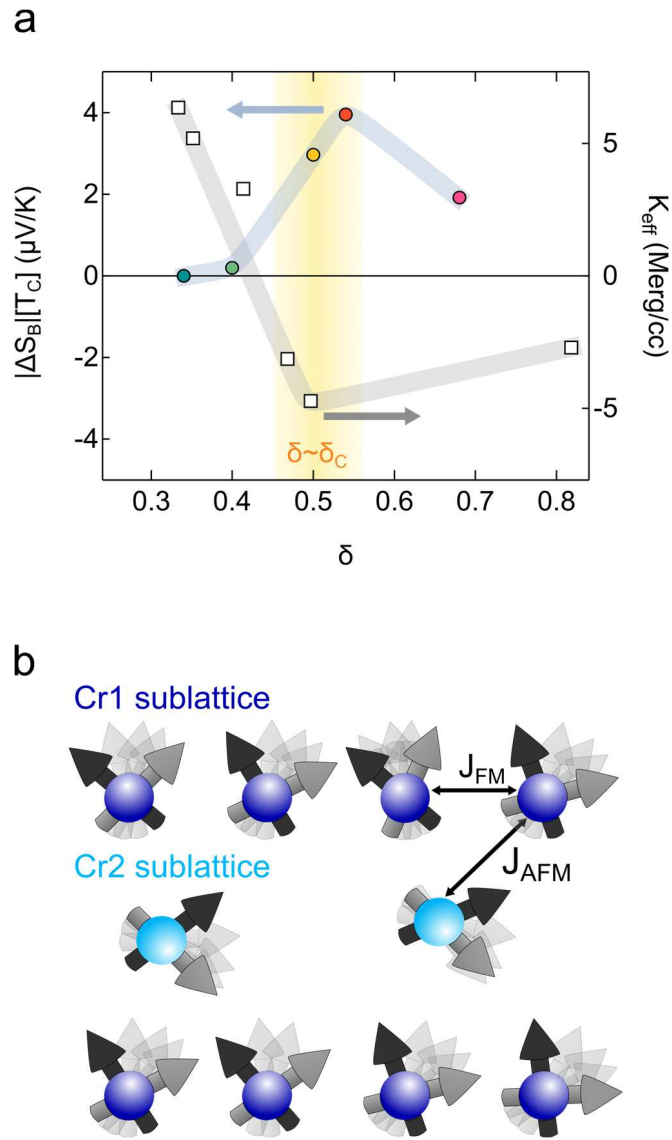

**Fig. 4** Existence of critical doping with AFM-enhanced magnetic fluctuations. **a** The phenomenological correspondence between the doping evolution of $|\Delta S_B(T_C)|$ (left axis) and that of $K_{eff}$ (right axis). Shaded regions are meant simply as a guide to the eye for critical region (see main body for details). **b** The schematic for magnetic competition and enhanced fluctuation around critical doping.